\documentclass[english, twocolumn, 10pt, aps, superscriptaddress, floatfix, showpacs, prb, citeautoscript]{revtex4-1}
\pdfoutput=1
\usepackage[utf8]{inputenc}
\usepackage[T1]{fontenc}
\usepackage{verbatim}
\usepackage{units}
\usepackage{mathtools}
\usepackage{amsmath}
\usepackage{amssymb}
\usepackage{graphicx}
\usepackage{wasysym}
\usepackage{layouts}
\usepackage{siunitx}
\usepackage{bm}
\usepackage{xcolor}
\usepackage[colorlinks, citecolor={blue!50!black}, urlcolor={blue!50!black}, linkcolor={red!50!black}]{hyperref}
\usepackage{bookmark}
\usepackage{tabularx}
\usepackage{microtype}
\usepackage{babel}
\hypersetup{pdfauthor={B. Nijholt and A. R. Akhmerov},pdftitle={Orbital effect of magnetic field on the Majorana phase diagram}}

\DeclareMathOperator{\Pf}{Pf}

\renewcommand{\comment}[2]{#2}

\DeclarePairedDelimiter\abs{\lvert}{\rvert}
\DeclarePairedDelimiter\norm{\lVert}{\rVert}

\makeatletter
\let\oldabs\abs
\def\abs{\@ifstar{\oldabs}{\oldabs*}}
\let\oldnorm\norm
\def\norm{\@ifstar{\oldnorm}{\oldnorm*}}
\makeatother

\newcolumntype{L}[1]{>{\raggedright\arraybackslash}p{#1}}
\newcolumntype{C}[1]{>{\centering\arraybackslash}p{#1}}
\newcolumntype{R}[1]{>{\raggedleft\arraybackslash}p{#1}}

\begin{document}

\title{Orbital effect of magnetic field on the Majorana phase diagram}
\author{Bas Nijholt}
\affiliation{Kavli Institute of Nanoscience, Delft University of Technology, P.O. Box 4056, 2600 GA Delft, The Netherlands}
\author{Anton R. Akhmerov}
\affiliation{Kavli Institute of Nanoscience, Delft University of Technology, P.O. Box 4056, 2600 GA Delft, The Netherlands}
\date{28 June 2016}

\begin{abstract}
Studies of Majorana bound states in semiconducting nanowires frequently neglect the orbital effect of a magnetic field.
Systematically studying its role leads us to several conclusions for designing Majoranas in this system. 
Specifically, we show that for experimentally relevant parameter values the orbital effect of a magnetic field has a stronger impact on the dispersion relation than the Zeeman effect.
While Majoranas do not require the presence of only one dispersion subband, we observe that the size of the Majoranas becomes unpractically large, and the band gap unpractically small, when more than one subband is filled.
Since the orbital effect of a magnetic field breaks several symmetries of the Hamiltonian, it leads to the appearance of large regions in parameter space with no band gap whenever the magnetic field is not aligned with the wire axis.
The reflection symmetry of the Hamiltonian with respect to the plane perpendicular to the wire axis guarantees that the wire stays gapped in the topologically nontrivial region as long as the field is aligned with the wire.
\end{abstract}

\maketitle

\section{Introduction}
The search for Majorana bound states, the simplest non-Abelian particles, is fueled by their suitability for fault-tolerant quantum computation~\cite{alicea_new_2012, beenakker_search_2013}.
A large fraction of the experimental effort~\cite{mourik_signatures_2012,das_zero-bias_2012,deng_anomalous_2012,churchill_superconductor-nanowire_2013,deng_parity_2014} is focused on creating Majoranas in semiconducting nanowires with proximity superconductivity, spin-orbit coupling, and magnetic field.
The theoretical foundation for this platform was initially developed for a single one-dimensional spinful band with intrinsic superconducting pairing~\cite{lutchyn_majorana_2010,oreg_helical_2010}.
Due to its compactness this model can be solved analytically, and it predicts that Majorana bound states appear when $E_\textrm{Z}^{2}>\mu^{2}+\Delta^{2}$, when the Zeeman energy becomes larger than the harmonic mean of the superconducting gap and the chemical potential.

\comment{Model is too simple and generalizations have been made.}
The single-mode model is minimalistic and neglects many physical phenomena that are crucial for understanding the properties of the Majorana bound states.
The existing extensions of this model study multimode wires~\cite{potter_multichannel_2010}, better modeling of the induced gap~\cite{liu_zero-bias_2012,stanescu_soft_2014}, the role of electrostatics~\cite{vuik}, disorder~\cite{potter_topological_2012,pientka_enhanced_2012,adagideli_effects_2014}, and the $k \cdot p$-model~\cite{stanescu_majorana_2013}.
The orbital effect of a magnetic field was analyzed both in planar wires~\cite{osca_majorana_2015, lim_magnetic-field_2012} and on the surface of a cylinder~\cite{lim_emergence_2013}.

\comment{We include orbital effects.}
We systematically study the influence of the orbital effect of a magnetic field on the symmetries of the Hamiltonian and the topological phase diagram for a three-dimensional (3D) nanowire.
The orbital effect of a magnetic field perpendicular to the wire induces a skipping orbit motion of the electrons.
The cyclotron radius becomes comparable to the typical wire diameters $d \sim \SI{100}{\nano\metre}$ already at the field of $\SI{0.3}{\tesla}$, and at chemical potential corresponding to the optimal topological band gap.
In addition, a field parallel to the wire shifts the energies of each band due to the effect of magnetic flux.
We expect the shift of the energies to be comparable to the level spacing when the flux through the wire diameter is of the order of a flux quantum.
Our findings are very different from those of Refs.~\onlinecite{osca_majorana_2015, lim_emergence_2013, lim_magnetic-field_2012} because we do not limit our analysis to a Hamiltonian with an artificially high spatial symmetry, or low dimensionality.

\section{Model}

\comment{We consider a three-dimensional nanowire with a Rashba-BdG Hamiltonian.}

We consider a 3D semiconducting nanowire with Rashba spin-orbit coupling and proximity-induced s-wave superconductivity.
The nanowire cross section is a regular hexagon, and the nanowire is translationally invariant in the $x$-direction.
Its Bogoliubov-de Gennes Hamiltonian (BdG) is
\begin{eqnarray}
H_\textrm{BdG} & = & \left(\frac{\bm{p}^{2}}{2m^*}-\mu\right)\tau_z+\alpha\left(p_{y}\sigma_x -p_{x}\sigma_y \right)\tau_z\nonumber \\
 &  & +\frac{1}{2}g\mu_\textrm{B}\bm{B}\cdot\boldsymbol{\sigma}+\Delta\tau_x,\label{eq:H_BdG}
\end{eqnarray}
and it acts on the spinor wave function $\Psi={\left(\psi_{e\uparrow},\psi_{e\downarrow},\psi_{\textrm{h}\downarrow},-\psi_{\textrm{h}\uparrow}\right)}^{T}$, where $\psi_e$, $\psi_\textrm{h}$ are its electron and hole components, and $\psi_\uparrow$, $\psi_\downarrow$ are the spin-up and spin-down components.
We introduced the Pauli matrices $\sigma_{i}$ acting on the spin degree of freedom and $\tau_{i}$ acting on the electron-hole degree of freedom.
Further $\bm{p}=-i\hbar\nabla+e\bm{A}\tau_z$ is the canonical momentum, with $e$ the electron charge and the vector potential $\bm{A}={\left[ B_y (z - z_0) - B_z (y - y_0), 0, B_x (y - y_0)\right]}^{T}$ chosen such that it does not depend on $x$.
We set the offsets $y_0$ and $z_0$ to ensure that the average vector potential vanishes in the superconductor.
This choice corresponds to a limit when the superconductor is thinner than the screening length and its total supercurrent is zero, appropriate for existing devices.
Finally, $m^*$ is the effective electron mass, $E_\textrm{Z}=\mu_\textrm{B}g\bm{B}\cdot\boldsymbol{\sigma}/2$ the Zeeman energy,  $\Delta$ the superconducting pairing potential, $\alpha$ the Rashba spin-orbit coupling strength, and $\mu=\mu_0+\mathcal{E} z$ the chemical potential created by a constant electric field $\mathcal{E}$ in the sample parallel to the $z$\kern-.05ex-axis, such that the Rashba spin-orbit acts in the $xy$\kern-.05ex-plane.

\comment{The pairing is either intrinsic or attached laterally.}

First we consider a model with a constant superconducting gap $\Delta$ inside the wire [see Fig.~\ref{fig:geometry}(a)] and then proceed to make a more realistic model of the superconductor.
To do that we set the superconducting order parameter $\Delta$ to zero in the wire and add a superconductor to the top which covers $3/8$ of the circumference of the wire [see Fig.~\ref{fig:geometry}(b)].
We choose the thickness of the superconductor to be $\SI{20}{\nano\metre}$ and set $\Delta$ in the superconductor such that the induced gap of the lowest band is $\Delta_\textrm{ind}=\SI{0.250}{\milli\electronvolt}$.
This is done by computing band energies at $k=0$ over a range of $\mu$ and matching the minimum to $\Delta_\textrm{ind}$.
We add a tunnel barrier between the two materials to change the transparency of the superconductor.
In the setup of Fig.~\ref{fig:geometry}(c), we break the reflection symmetry with respect to the $xz$-plane by moving the superconductor to the side similar to the experimental setup of Mourik \emph{et al}~\cite{mourik_signatures_2012}.

\comment{We discretize the model and simulate it with Kwant.}

To perform the numerical simulations we discretize the Hamiltonian on a cubic lattice with lattice constant $a=\SI{10}{\nano\metre}$, much smaller than the minimal Fermi wavelength in the parameter range we consider.
The discretization does not break or introduce any additional symmetries.
The Hamiltonian at a lattice momentum $k$ equals $H\left(k\right)=h+t\exp(ik)+t^{\dagger}\exp(-ik)$ where $h$ is the Hamiltonian of the cross section of the tight-binding system and $t$ is the hopping matrix between the neighboring cross sections.
We introduce the vector potential by Peierls substitution $t_{nm}\rightarrow t_{nm}\exp(-ie\intop\bm{A}d\bm{l})$~\cite{hofstadter_energy_1976}.
We perform the numerical simulations using the Kwant code~\cite{groth_kwant:_2014}.
The source code and the specific parameter values are available in the Supplemental Material~\cite{supp}. 
The resulting raw data are available in Ref.~\onlinecite{data}.

\section{Symmetry analysis}

\comment{The fundamental symmetry of the Majoranas is class D, but it does not allow reliable creation of topological gap because it allows for the tilting of the bands.}

The Majorana bound states are protected by the combination of the band gap and the particle-hole symmetry of the Hamiltonian $\mathcal{P}H\left(k\right)\mathcal{P}^{-1}=-H\left(-k\right)$.
In the basis of Eq.~\eqref{eq:H_BdG} this symmetry has the form $\mathcal{P}=\sigma_y \tau_y\mathcal{K}$, with $\mathcal{K}$ the complex conjugation.
In general there are no additional symmetries and the Hamiltonian belongs to symmetry class D~\cite{altland_nonstandard_1997-1}.
Particle-hole symmetry only requires that the energy $E_n(k)$ of $n$-th band at momentum $k$ is $E_n(k)=-E_m(-k)$ of some other $m$-th band; at the same time $\mathcal{P}$ puts no constraints on $E_n$ itself.
This means that whenever $E_{n}$ changes sign at a certain momentum, the band structure becomes gapless.
This tilting of the band structure~\cite{rex2014tilting} [shown in the middle panels in Fig.~\ref{fig:bandstructure}, where $E_n(k) \ne -E_m(k)$] is a strong effect that does not vanish with superconducting gap or spin-orbit, and can easily become larger than the induced gap, rendering the creation of Majoranas impossible.

\comment{In the existing models have an extra chiral symmetry that prevents tilting, but it is broken by the orbital part of magnetic field.}

The tilting of the band structure is absent if the Hamiltonian has an extra chiral symmetry alongside $\mathcal{P}$.
It has been shown that the Hamiltonian has an approximate chiral symmetry $\mathcal{C}H\left(k\right)\mathcal{C}^{-1}=-H\left(k\right)$, $\mathcal{C}=\sigma_y \tau_y$, valid when the wire diameter $d$ is smaller than the spin-orbit length $l_{so}=\hbar^{2}/m^{*}\alpha$~\cite{tewari_topological_2012,diez_andreev_2012}, and $B_y=0$.
Then the $p_{y}\sigma_x \tau_z$ term, associated with the transverse motion in Eq.~\eqref{eq:H_BdG}, is negligible.
Without the tilting, the system is gapped in every region of parameter space, except at the topological phase boundaries.
However, for relevant experimental parameters~\cite{mourik_signatures_2012}, the orbital terms break this symmetry more strongly than the spin-orbit term, bringing the system back to symmetry class D.

\comment{We find that there is a reflection symmetry that guarantees the absence of tilting for field in $x$.}

We perform a systematic search of symmetries that the Hamiltonian~\eqref{eq:H_BdG} may have.~\cite{rosdahl}
We find the reflection symmetry with respect to the $yz$-plane $\mathcal{R}_x H\left(k\right)\mathcal{R}_x^{-1}=H\left(-k\right)$, $\mathcal{R}_x=\sigma_x\delta(x+x')$.
It is independent of the wire geometry and spin-orbit strength and guarantees the absence of tilting whenever the field is aligned with the $x$-axis.
The combined symmetry $\mathcal{P}' = \mathcal{R}_x \mathcal{P}$ is local in momentum space and ensures the absence of band structure tilting: $\mathcal{P}' H(k) \mathcal{P}'^{-1} = -H(k)$.

\comment{Additionally, if the system has a reflection symmetry along the $y$-axis, there is an extra chiral symmetry that is valid for fields in the complete $xz$-plane.}

\begin{figure}
\includegraphics[width=0.95\columnwidth]{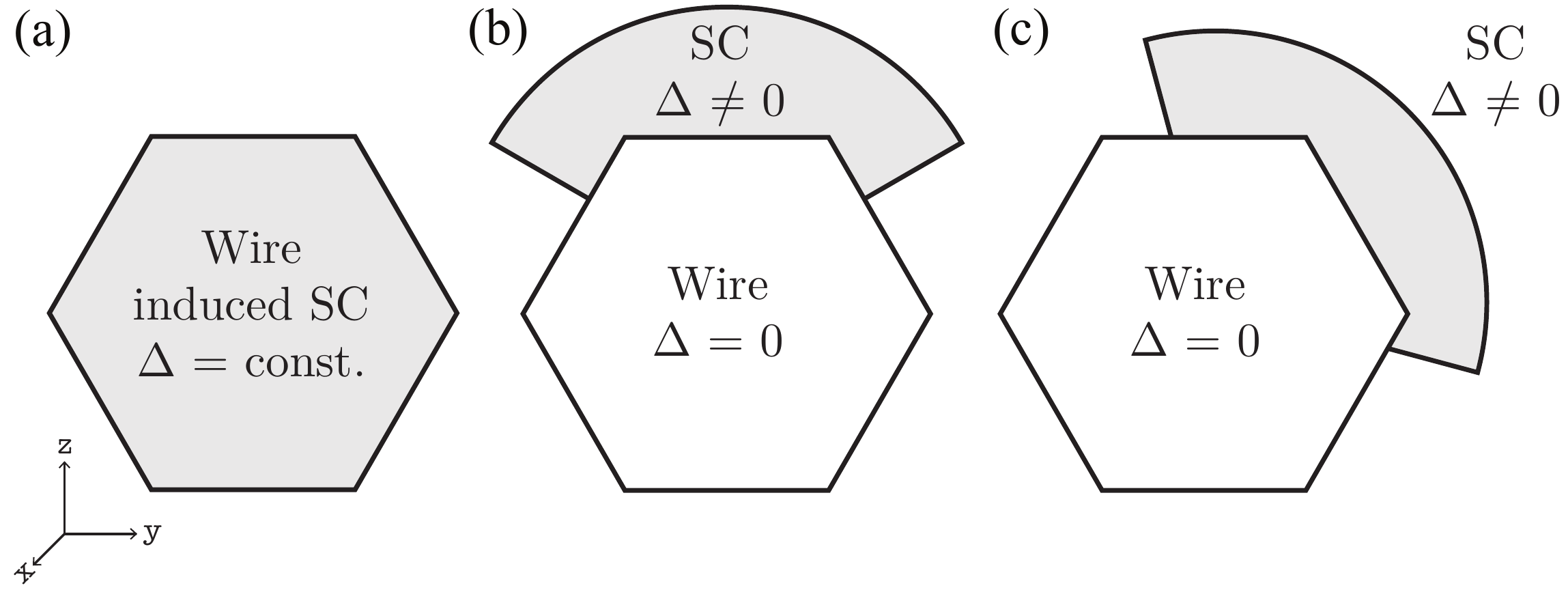}
\caption{Three hexagonal nanowire devices we consider: (a) (a) with an intrinsic pairing term and with a proximity-coupled superconductor (b) on the top and (c) on the side.
The last two setups have tunnel barriers between the superconductor and the nanowire.
\label{fig:geometry}}
\end{figure}

\begin{figure}
\includegraphics[width=0.95\columnwidth]{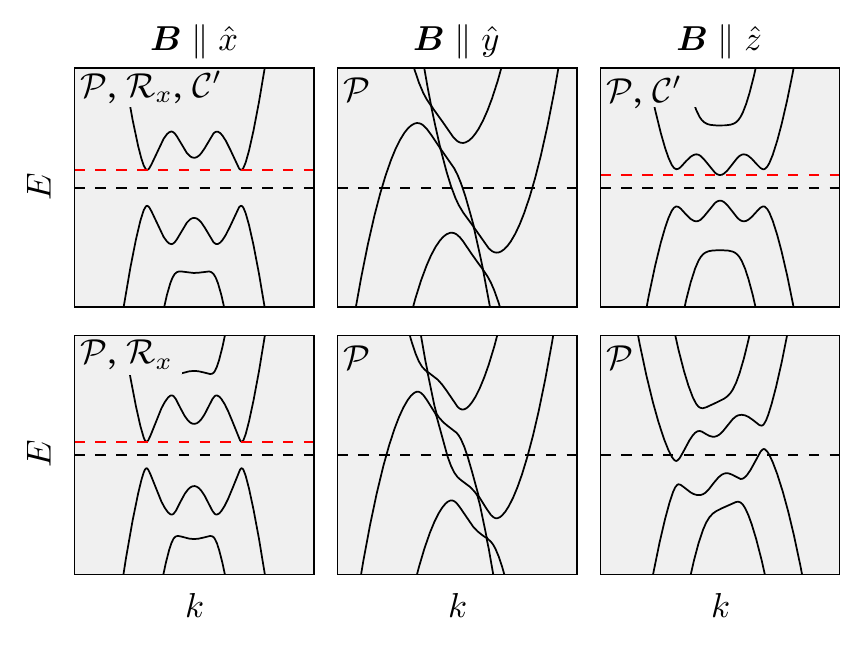}
\caption{Band structures of the setup of Fig.~\ref{fig:geometry}(b) (top) and Fig.~\ref{fig:geometry}(c) (bottom).
Each panel is labeled with the symmetries respected by the corresponding Hamiltonian.
The dashed black line indicates the Fermi energy $(E=0)$.
The red dashed lines show the size of the band gap if it is present.
In the top row, the reflection symmetry of the wire along the $y$-axis $\mathcal{R}_y$ makes the Hamiltonian have a chiral symmetry $\mathcal{C}'$ when the magnetic field lies in the $xz$-plane.
The wire used for the calculation of the bottom row dispersions lacks $\mathcal{R}_y$ and therefore has no $\mathcal{C}'$.
Without $\mathcal{C}'$ the bands are allowed to tilt and the gap may close whenever $B_y \neq 0$ or $B_z \neq 0$.
A magnetic field parallel to the $x$-axis preserves $\mathcal{R}_x$, which protects the band gap from closing.\label{fig:bandstructure}}
\end{figure}

Additionally, we find a chiral symmetry $\mathcal{C}'=\tau_y\mathcal{R}_y$, $\mathcal{C}'H\left(k\right)\mathcal{C}'^{-1}=-H\left(k\right)$, with $\mathcal{R}_y = \sigma_y \delta(y + y')$ the reflection with respect to the $y$-axis.
This chiral symmetry holds when the magnetic field lies in the $xz$-plane and none of the potentials in Eq.~\eqref{eq:H_BdG} break $\mathcal{R}_y$, like in the setups of Figs.~\ref{fig:geometry}(a) and\ref{fig:geometry}(b).
When present, $\mathcal{C}'$ guarantees the absence of band structure tilting just like $\mathcal{C}$.
This symmetry is present in most theoretical models, and in particular it is obeyed by the Hamiltonians used in Refs.~\onlinecite{osca_majorana_2015, lim_emergence_2013, lim_magnetic-field_2012}.
A finite $B_y$ breaks both $\mathcal{R}_x$ and $\mathcal{C}'$ therefore, the bands can tilt and close the topological gap.
The band structures in Fig.~\ref{fig:bandstructure} summarize the relation between the geometry of the setup of Figs~\ref{fig:geometry}(b) and Fig.~\ref{fig:geometry}(c), magnetic field orientation, and the symmetries of the Hamiltonian.

\section{Calculating the topological phase diagram}

\comment{We use a generalized eigenvalue problem to find all phase boundaries at once.}

\begin{figure}
\includegraphics[width=0.95\columnwidth]{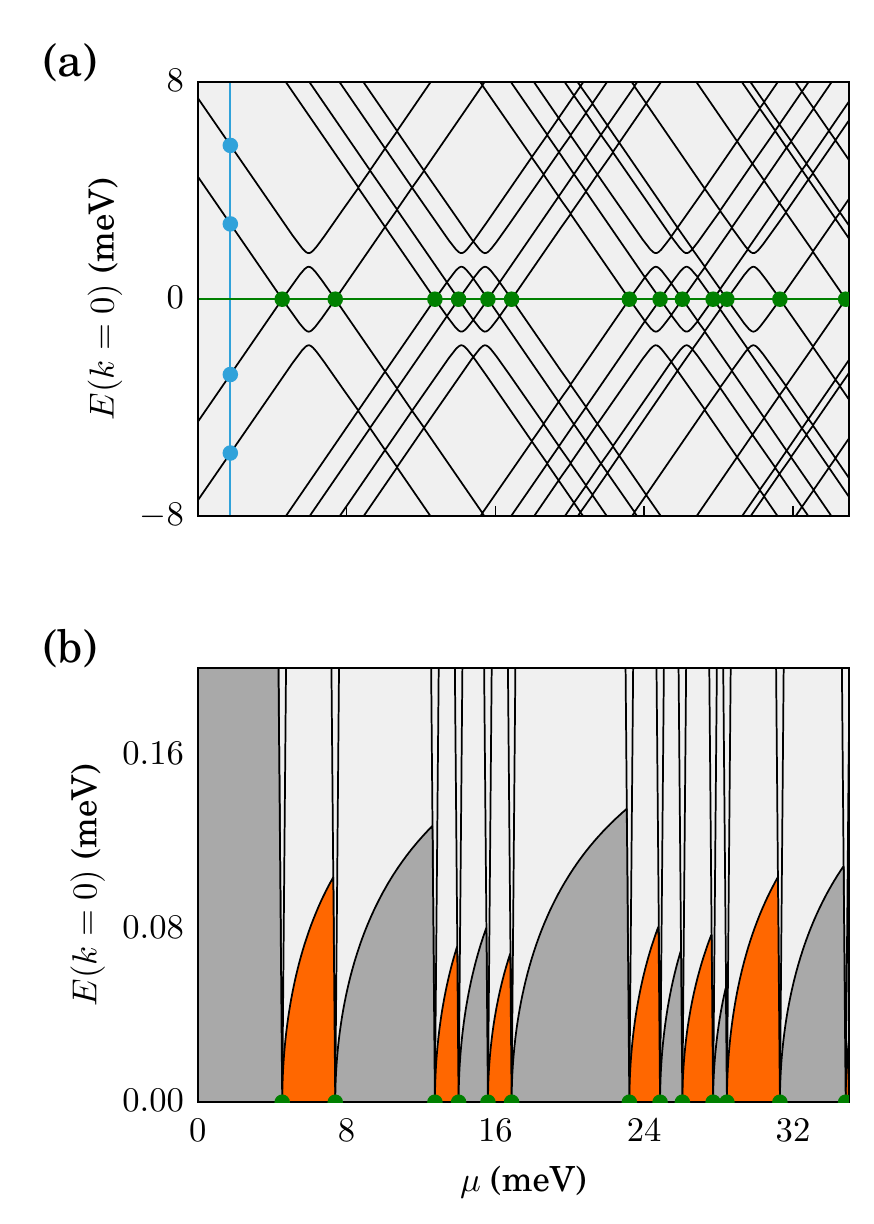}
\caption{(a) Energy spectrum at $k=0$ of the setup of Fig.\ref{fig:geometry}(a) as a function of chemical potential $\mu$.
The blue points are the solutions of $H_\textrm{BdG} \psi = E \psi$ at fixed $\mu$ marked by the blue line.
The green points are the real eigenvalues of Eq.~\eqref{eq:eigenproblem} lying at $E=0$ (the green line).
(b) The gap size for the same setup and parameters, with dark gray regions trivial and the orange regions topological.\label{fig:ev_problem}}
\end{figure}

We use an optimized algorithm to quickly find all the $\mu$ values corresponding to the topological phase transitions at once.
The topological transitions in symmetry class D occur when $\Pf H_\textrm{BdG}(k=0)$ changes sign.
\footnote{The band gap closings at $k=\pi$ can be treated identically, but they never appear in our model Hamiltonian.}
Since the sign change of $\Pf H_\textrm{BdG}$ is accompanied by the appearance of zero energy states, we need to find $\mu$ and $\psi$ such that $H(\mu, k=0)\psi = 0$.
Using that $\mu$ enters the Hamiltonian as a prefactor of a linear operator, we rewrite this equation as a generalized eigenproblem:
\begin{eqnarray}
H_\textrm{BdG}\left(\mu=0,k=0\right)\psi=\mu\tau_z\psi.\label{eq:eigenproblem}
\end{eqnarray}
The real eigenvalues of this eigenproblem are the values of $\mu$ where the gap closes at $k=0$ [see Fig.~\ref{fig:ev_problem}(a)], and they can be found using standard generalized eigensolvers.
If the dispersion relation is gapped also at any finite $k$, these gap closings are the boundaries of the topological phase.

Since the eigenvalues of $H_\textrm{BdG}$ come in opposite sign pairs, the real eigenvalues of Eq.~\eqref{eq:eigenproblem} always come in degenerate pairs, and each pair lies at a transition between a trivial and a nontrivial phase.
We complete the calculation of the topological phase diagram by using as a reference point that $H_\textrm{BdG}(\mu = -\infty)$ is topologically trivial.

\begin{figure}
\includegraphics[width=0.95\columnwidth]{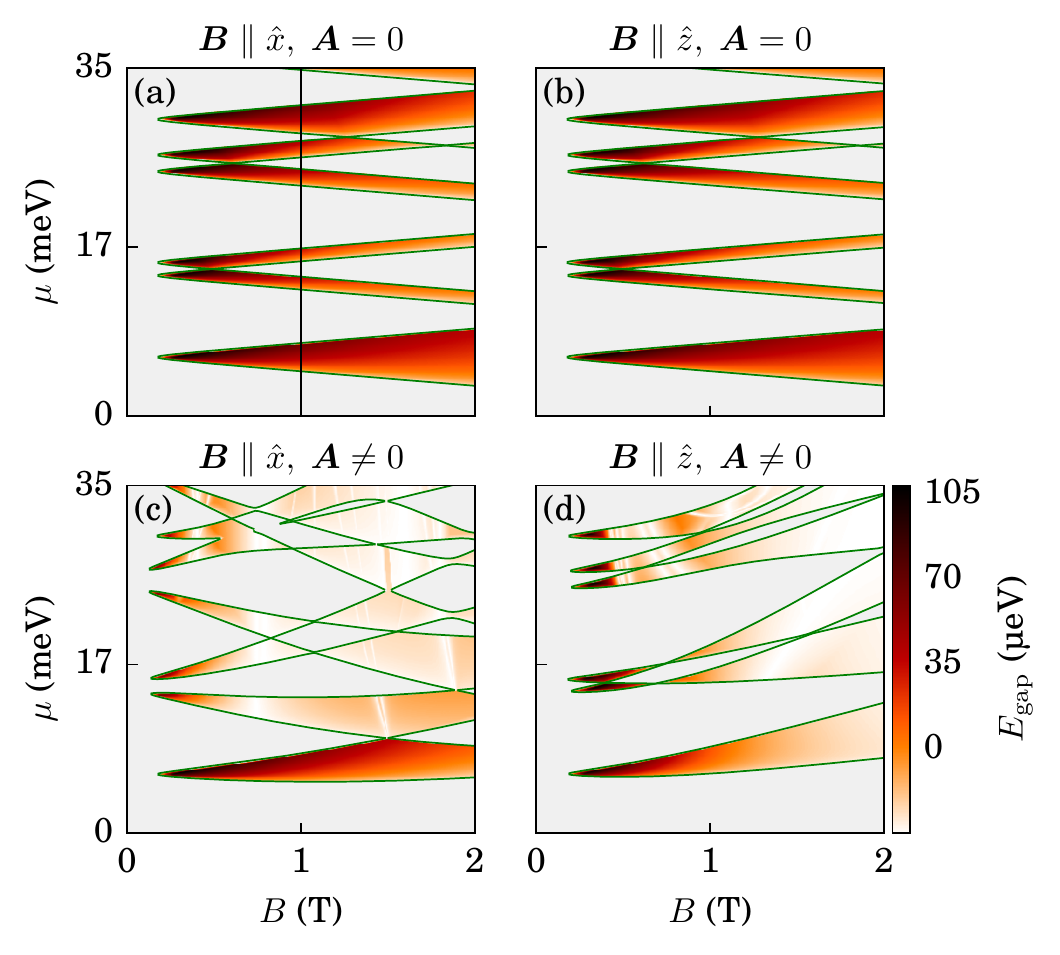}
\caption{Phase diagrams of the setup of Fig.~\ref{fig:geometry}(a) (a), (b) without
the orbital effect of a magnetic field and (c), (d) with it.
The green lines depict the topological phase transitions.
The colored regions are topologically nontrivial, with the color representing the size of the topological band gap $E_\textrm{gap}$.
At $B \gtrsim \SI{1}{\tesla}$ the orbital effect of the magnetic field becomes stronger than the Zeeman effect and changes the sign of the slope of half of the phase boundaries.
Furthermore, the orbital effect leads to a faster suppression of the band gaps with magnetic field.\label{fig:phase-diagrams}
The narrow regions with suppressed $E_\textrm{gap}$ originating from the crossings of the phase boundaries in (c) are due to Dirac cones appearing in $(k_x, B)$\!-space and are protected by $C'$.
The vertical black line in (a) indicates the value of the magnetic field used in Fig.~\ref{fig:ev_problem}.}
\end{figure}

\comment{We calculate the gap by performing a binary search for the energy at which propagating modes appear.}

The generalized eigenvalue algorithm for finding phase boundaries does not guarantee that $H(k)$ is gapped for $k \neq 0$, and therefore we calculate the magnitude of the gap $E_\textrm{gap}$ in the topologically nontrivial regime separately for each set of parameter values.
We form a translation eigenvalue problem to calculate all the modes of $H_\textrm{BdG}$ at a given energy $E$ and check whether there are any propagating modes~\cite{groth_kwant:_2014}.
By using a binary search in $E$ for the energy at which the propagating modes start to appear, we find $E_\textrm{gap}$ [see Fig.~\ref{fig:ev_problem}(b)].

\comment{We find the Majorana lengths by calculating the slowest decaying mode.}

The real space size of the Majoranas $\xi$ imposes a lower bound on the nanowire length required to create them.
To calculate $\xi$ we find the eigenvalue decomposition of the translation operator at zero energy.
The eigenvalue $\lambda_{\min}$ closest to the unit circle corresponds to the slowest decaying part of the Majorana wave function.
We calculate $\xi$ using
\begin{equation}
\xi=\abs{\log^{-1}\abs{\lambda_{\min}}}.
\end{equation}

\begin{figure}
\includegraphics[width=0.95\columnwidth]{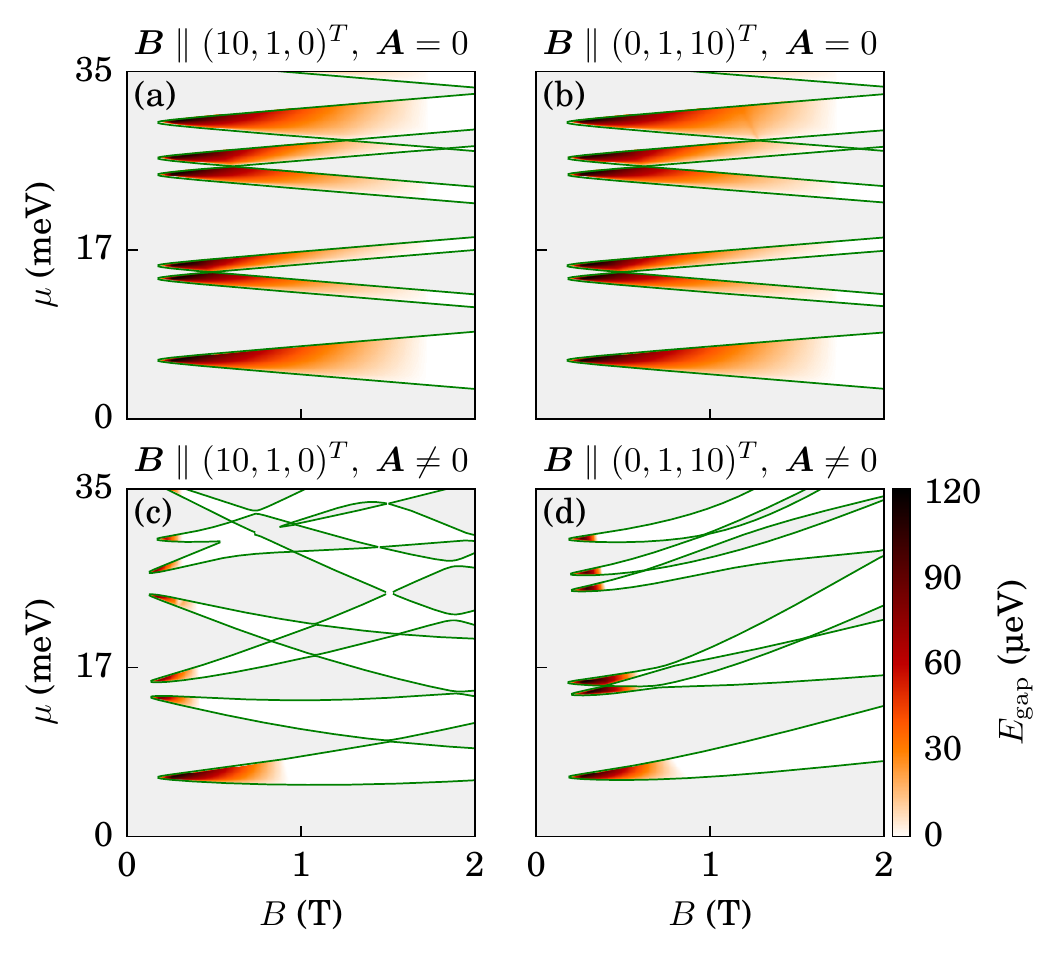}
\caption{Same as Fig.~\ref{fig:phase-diagrams}, but with the magnetic field slightly misaligned.
We observe that the band gaps close quickly upon changing the direction of the magnetic field towards the spin-orbit direction in $y$.
\label{fig:misaligned}}
\end{figure}

\begin{figure}
\includegraphics[width=0.95\columnwidth]{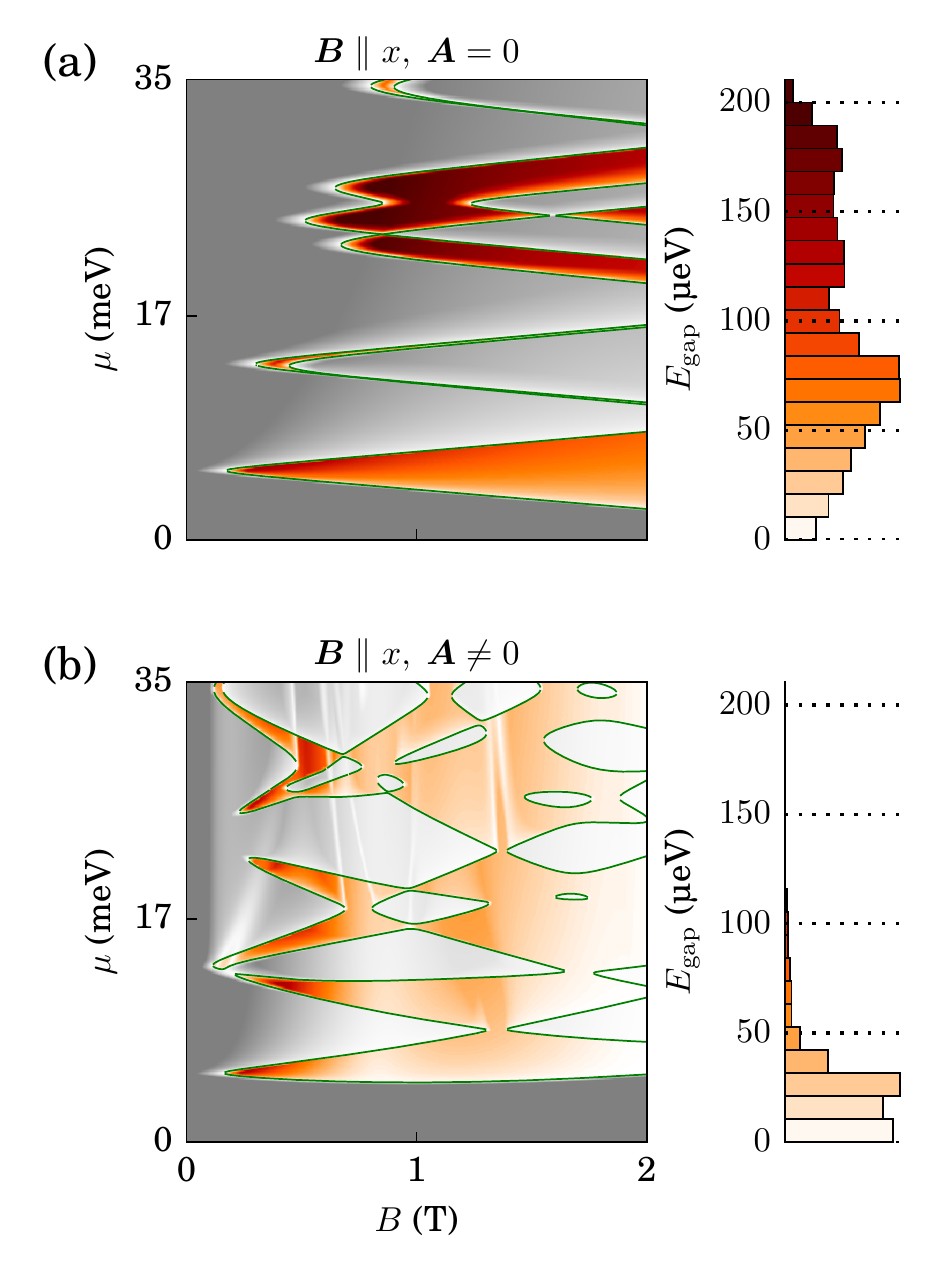}
\caption{Phase diagrams of the setup of Fig.~\ref{fig:geometry}(c) (a) without the orbital effect of a magnetic field and (b) with it. Color scale corresponds to $E_\textrm{gap}$, with the topological regions colored and trivial regions in grayscale.
The histograms in the right-hand panels show the distribution of the gap values sampled in the topological regime within the selected parameter range. 
Neglecting the orbital effect of the magnetic field incorrectly leads to a strong dependence of the critical field on $\mu$.
With the orbital effect of magnetic field flux penetration through the quasiparticle trajectory changes the interference phases and can suppress the topological gap $E_\textrm{gap}$.
\label{fig:phase-diagram-sc_on_side}}
\end{figure}

\begin{figure}
\includegraphics[width=0.95\columnwidth]{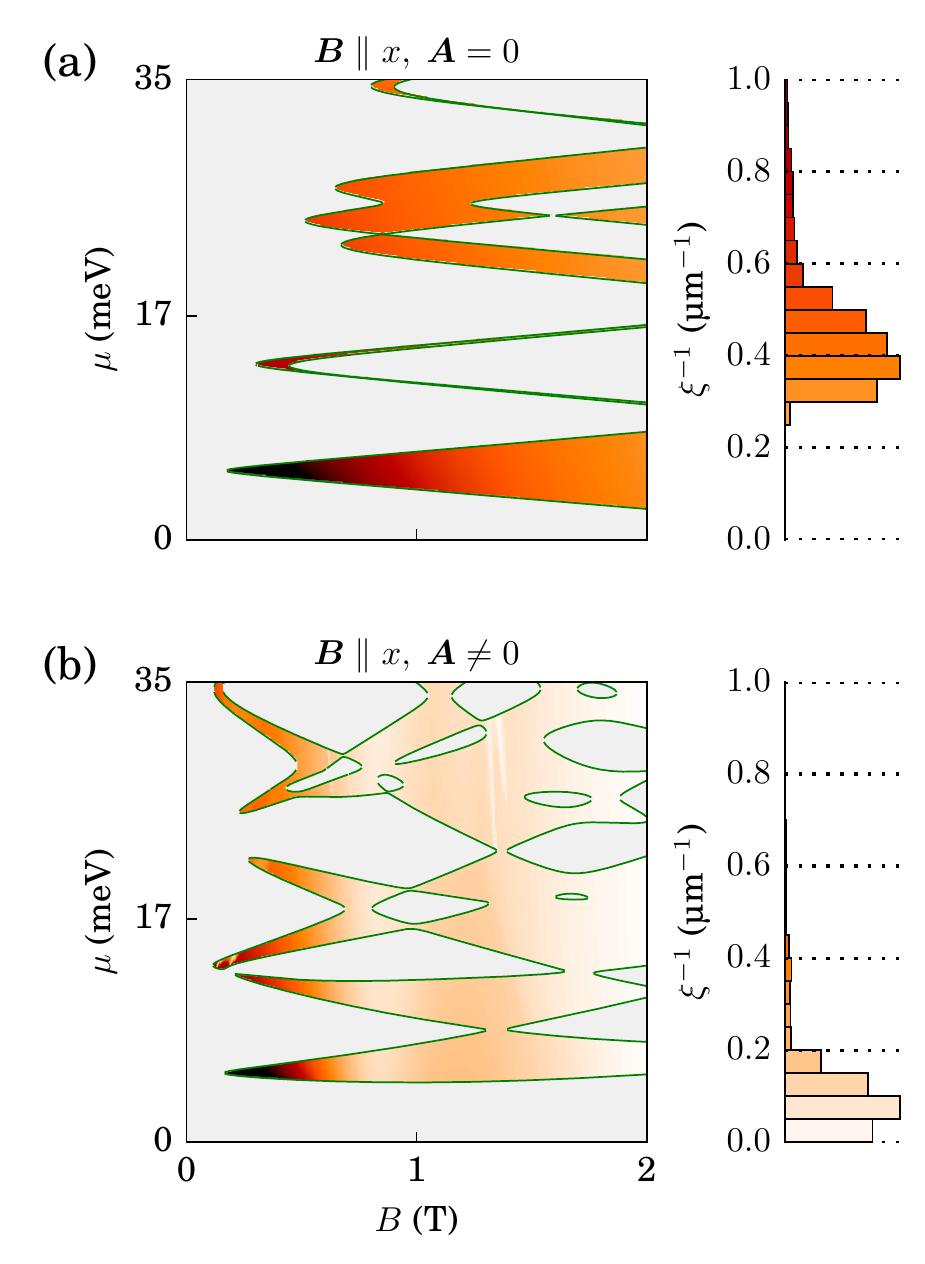}
\caption{Same as Fig.~\ref{fig:phase-diagram-sc_on_side}, but with color representing inverse Majorana length $\xi^{-1}$.
The histogram and color scales are truncated from above at $\SI{1}{\per\micro\metre}$.
The mode of the distribution of $\xi^{-1}$ reduces from $\SI{0.35}{\per\micro\metre}$ to $\SI{0.10}{\per\micro\metre}$ upon taking the orbital effect into account.
Although the Majorana lengths are overall much larger with the orbital effect of the magnetic field, the minimal length is close to $\SI{200}{\nano\metre}$ in both cases.\label{fig:Majorana-length}}
\end{figure}

\section{Results}

\comment{We find that for relevant experimental parameters~\cite{mourik_signatures_2012}, Zeeman field effect is weaker than orbital.}

We use realistic parameters of an InSb nanowire~\cite{mourik_signatures_2012}: $\alpha=\SI{20}{\milli\electronvolt \nano\metre}$, $m^{*}=0.015m_e$, $\Delta=\SI{0.250}{\milli\electronvolt}$, $d=\SI{100}{\nano\metre}$, and $g=50$.
At the high fields that are typically used in experiment ($B\apprge\SI{1}{\tesla}$), we find that the Zeeman effect of the magnetic field has a lower impact on the phase boundaries than the orbital effect of the magnetic field (see Fig.~\ref{fig:phase-diagrams}).
We verify that the band gap is protected by $\mathcal{C}'$ as long as $B_y = 0$, despite that the orbital effect of the magnetic field reduces $E_\textrm{gap}$.

In agreement with our expectations a finite $B_y \lesssim \SI{0.1}{\tesla}$ leads to the closing of the band gap (see Figs.~\ref{fig:bandstructure} and \ref{fig:misaligned}).
The maximum tolerable $B_y $ becomes smaller with increasing $\mu$.
The narrow regions with suppressed $E_\textrm{gap}$ visible in Figs.~\ref{fig:phase-diagrams}(c) and \ref{fig:phase-diagrams}(d) are the consequence of Dirac cones appearing in $(k_{x}, B)$-space and are protected by $\mathcal{C}'$.
Breaking $\mathcal{R}_y$ breaks $\mathcal{C}'$ and removes these Dirac cones.

\comment{With the reflection symmetry the gap closes quickly for fields in $y$ and $z$.}

We now turn to study the system shown in Fig.~\ref{fig:geometry}(c) that has $\mathcal{C}'$ strongly broken and only $\mathcal{R}_x$ and $\mathcal{P}$ remaining.
Since the induced superconducting gap $\Delta_\textrm{ind}\approx \SI{250}{\micro\electronvolt}$ in Ref.~\onlinecite{mourik_signatures_2012} is much smaller than the $\mathrm{NbTiN}$ gap $\SI{2}{\milli\electronvolt}$, the system must be in the long junction limit, where $E_\textrm{Th} \ll \Delta$.
In the long junction limit the induced gap equals $\Delta_\textrm{ind}\approx\hbar T v_\textrm{F} / d$, where $T$ is the transparency of the tunnel barrier, and $v_\textrm{F}$ the Fermi velocity.
In the absence of the orbital effect of a magnetic field, this means that the Zeeman energy has to exceed $\Delta_\textrm{ind}$ and therefore the critical value of the magnetic field at which the gap closes strongly depends on $\mu$ as seen in Fig.~\ref{fig:phase-diagram-sc_on_side}(a).
With the orbital effect of the magnetic field flux, penetration through the quasiparticle trajectory changes the interference phases, which suppresses the induced gap and causes the topological phase transitions to occur at a value of $B$ corresponding to a single flux quantum penetrating the wire area [see Fig.~\ref{fig:phase-diagram-sc_on_side}(b)].

The Majorana decay lengths $\xi$ significantly increase when including the orbital effect of the magnetic field in the Hamiltonian (see Fig.~\ref{fig:Majorana-length}).
Specifically, the mode of the distribution of $\xi$ changes by a factor of $\sim 4$ in the parameter range we consider (see histograms in Fig.~\ref{fig:Majorana-length}).
However, the minimum values of $\xi$ without orbital effect and with it are both $\approx \SI{200}{\nano\metre}$.
Therefore, $\mu$ needs to be tuned with sub-\unit{meV} precision within the lowest band in order to create Majorana bound states with practically relevant parameters.

\comment{Higher spin-orbit strength does not seem to matter.}
To investigate the effect of the spin-orbit coupling on the Majorana properties in the presence of an orbital field, we have repeated the calculations shown in Figs.~\ref{fig:phase-diagram-sc_on_side} and \ref{fig:Majorana-length} using a fivefold larger spin-orbit strength reported in Ref.~\onlinecite{Weperen2014}.
We find that the topological band gap increases overall and in particular the maximal gap grows from $\SI{0.14}{\milli\electronvolt}$ to $\SI{0.21}{\milli\electronvolt}$, while the minimal decay length remains almost the same.
Therefore, increasing spin-orbit strength has a positive but not very strong effect on the topological band gap.

\section{Discussion and Conclusions}

\comment{Our results: orbital field makes everything harder}
We have shown that the orbital effect of a magnetic field complicates the creation of Majoranas in nanowires.
Orbital terms break the chiral symmetry $\mathcal{C}$ and prevent the appearance of Majoranas whenever the magnetic field is not aligned with the wire axis.
When the field does point along the $x$-axis, we find that the reflection symmetry $\mathcal{R}_x$ in combination with particle-hole symmetry $\mathcal{P}$ protects the band gap from closing everywhere in $(B,\mu)$-space, except at the topological phase boundaries.
At experimentally relevant values of magnetic field, the orbital effect has a stronger impact on the dispersion relation than the Zeeman effect.
Furthermore, the orbital effect suppresses $E_\textrm{gap}$ and increases $\xi$.
However, the maximum value of the $E_\textrm{gap}$ in the topologically nontrivial region does not change as drastically (from $\SI{0.21}{\milli\electronvolt}$ to $\SI{0.14}{\milli\electronvolt}$) and the minimal decay length changes even less (from $\SI{201}{\nano\metre}$ to $\SI{210}{\nano\metre}$).
The reflection symmetry $\mathcal{R}_x$ of the Hamiltonian that we consider is respected by any Rashba spin-orbit interaction.
Dresselhaus spin-orbit coupling breaks $\mathcal{R}_x$; however, it is expected to be weak in the nanowires.

\comment{Extensions: we omit several physical effects.}
Our simulations can be made more complete by complementing them with self-consistent electrostatics and magnetic field screening by the superconductor.
An additional extension of our work is to go beyond the effective mass approximation and to use the $k \cdot p$ model.
A separate topic of study is the interplay between the orbital effect of the magnetic field and disorder.
We expect that the sensitivity to disorder will increase by taking the orbital effect of the magnetic field into account.

\comment{Outlook: how to cope with the complications.}
Our results suggest that keeping the chemical potential low is required to obtain Majoranas with reasonable length and energy scales.
Furthermore, our findings reveal a complication in realizing more sophisticated Majorana setups, such as a T-junction required for braiding.
This is because of the requirement that the field should be aligned with the nanowire.
A possible strategy to reduce the undesirable orbital effect of the magnetic field is to use nanowires with smaller diameters at a cost of reduced electric field effect and increased disorder sensitivity.

\comment{Acknowledgments}

We are grateful to T. Ö. Rosdahl, S. Rubbert, D. Sticlet, and M. Wimmer for useful discussions and L. P. Kouwenhoven for his support.
This work was supported by the Netherlands Organization for Scientific Research (NWO/OCW), as part of the Frontiers of Nanoscience program, the Foundation for Fundamental Research on Matter (FOM), and an ERC Starting Grant STATOPINS 638760.

\bibliographystyle{apsrev4-1}
\bibliography{orbitalmajorana}
\end{document}